\documentclass{article}
\usepackage{varioref}
\usepackage{amssymb}
\usepackage{color}
\usepackage{graphicx}
\usepackage{color}
\begin{document}

\title{\bf  Vaidya Spacetime in Massive Gravity's Rainbow }
\author{ Yaghoub Heydarzade $^{1}$\thanks{email: heydarzade@azaruniv.edu}\, Prabir Rudra $^{2}$ \thanks{email: prudra.math@gmail.com}\,
Farhad Darabi $^{1}$ \thanks{email: f.darabi@azaruniv.edu; Corresponding author}\,
Ahmed Farag Ali $^{3,4}$ \thanks{email: ahmed.ali@fsc.bu.edu.eg}\,
Mir Faizal $^{5,6}$ \thanks{email: mir.faizal@ubc.ca; mir.faizal@uleth.ca}
\\{\small $^1$ \emph{Department of Physics, Azarbaijan Shahid Madani University, Tabriz, 53714-161, Iran}}\\
{\small $^2$\emph{Department of Mathematics, Asutosh College, Kolkata-700 026, India} }\\
{\small $^3$ \emph{Netherlands Institute for Advanced Study,
Korte Spinhuissteeg 3, 1012 CG Amsterdam, Netherlands}}\\
{\small $^4$ \emph{Department of Physics, Faculty of Science,
Benha University, Benha, 13518, Egypt}}\\
{\small $^5$\emph{Irving K. Barber School of Arts and Sciences, University of British Columbia-Okanagan, Kelowna, BC V1V 1V7, Canada}}\\
{\small $^6$\emph{Department of Physics and Astronomy, University of Lethbridge, Lethbridge, AB T1K 3M4, Canada}}
}
\date{\today}
\maketitle

\begin{abstract}
In this paper, we will analyze the energy dependent
deformation of massive gravity using the formalism of massive gravity's rainbow.
So, we will use the Vainshtein mechanism and the  dRGT mechanism for the energy dependent massive gravity,
and thus analyze a ghost free theory of massive gravity's rainbow.
We study the energy dependence of a time-dependent geometry,   by
analyzing the  radiating Vaidya  solution in this theory of   massive gravity's rainbow.
The energy dependent deformation  of this Vaidya metric will be  performed using suitable  rainbow functions.
\end{abstract}

\section{Introduction}
It is expected that the usual energy-momentum dispersion relation
will get deformed in the UV limit due to quantum gravitational
effects. In fact, such a deformation of the usual energy-momentum
dispersion relation has been observed to occur in
  loop quantum gravity  \cite{AmelinoCamelia:1996pj}-\cite{amerev}, discrete spacetime~\cite{'tHooft:1996uc},
 string field theory~\cite{LIstring},
spacetime foam~\cite{AmelinoCamelia:1997gz},  spin-networks~\cite{Gambini:1998it}, and
 non-commutative geometry~\cite{Carroll:2001ws}. As the usual energy-momentum dispersion relation is fixed by
 Lorentz symmetry, the deformation of the usual energy-momentum dispersion relation in the UV limit seems to indicate a breaking
 of Lorentz symmetry in the UV limit. In fact, such a violation of Lorentz symmetry can be used to explain
 anomalies in ultra-high energy cosmic rays and
TeV photons~\cite{ AmelinoCamelia:1997jx}-\cite{AmelinoCamelia:1999wk}.
It may be noted that the threshold anomalies are only predicted by  deformations where  the usual energy-momentum
dispersion relation is deformed by a preferred reference frame, and they  do not occur in   deformation where
 no such preferred reference frame exists~\cite{AmelinoCamelia:2002dx}.
 The deformation of the usual energy-momentum dispersion relation can be explained using the
 doubly special relativity (DSR) \cite{AmelinoCamelia:2000mn}-\cite{AmelinoCamelia:2000mn1}.
 The
DSR is an extension of the  special theory of relativity, in which the Planck energy and the velocity  of light are universal
constants. So, just as in special relativity, no object can attain a velocity greater than the velocity of light, in DSR
no object can have an energy greater than the Planck energy.
The DSR can be generalized to curved spacetime, and the resulting theory is called gravity's rainbow  \cite{Magueijo:2002xx}.
In this formalism, the spacetime geometry is described by a rainbow of energy dependent metrics, as the geometry
of spacetime depends on the energy of the probe.  The gravitational dynamics in gravity's rainbow can be studied using
rainbow functions~\cite{Galan:2004st}--\cite{Galan:2004st1}. The gravity's rainbow has been used to study
inflation~\cite{Barrow:2013gia}-\cite{Barrow:2013gia1},
and a resolving of the Big Bang singularity~\cite{Awad:2013nxa}-\cite{FRWRainbow1}.
 It may be noted that gravity's rainbow is related to Horava-Lifshitz gravity
 \cite{HoravaPRD}-\cite{HoravaPRL},
and for a specific choice of rainbow functions, it
produces the same results as produced by Horava-Lifshitz gravity \cite{re}.

 The  main motivation for gravity's rainbow comes from  the observation that the supergravity is a low energy approximation
 to the  string theory \cite{lowe1}-\cite{s1t2}.
  This is because according to
the renormalization group flow,  constants   depend on the scale at which a theory
\cite{renom}-\cite{renor}.
Furthermore, the scale at which a theory is measured will depend on the energy of
the probe used to measure such a
theory. Thus, as the constants in a theory depend explicitly on the scale at which
a theory is measured, they
also depend implicitly on the energy of the probe used to measure such constants.
Now string theory can be viewed as a
two dimensional theory, and
the target space metric can  be regarded as  a matrix of coupling constants of this two
dimensional theory.
As these coupling constants would flow and depend explicitly on the scale at which the
theory is measured,
they would   implicitly depend on the
energy of the probe used to perform such a measurement.
This would make the metric of spacetime depend on the energy of
the probe, and thus we would obtain gravity's rainbow  \cite{Magueijo:2002xx}.

So, the gravity's rainbow can be motivated from string theory, as the energy dependence of the
spacetime metric can be motivated from the flowing of target space geometry in string theory.
It may be also noted that  various solution obtained in string theory
have been generalized to massive gravity, which is a theory with massive gravitons.
In fact, massive Type IIA supergravity has been studied \cite{iiaa12}-\cite{iiaa14}.
The Fermionic T-duality has also been studied for  massive type IIA supergravity \cite{iiaa15}.
The relation between the massive  IIA supergravity and M-theory has also been investigated  \cite{iiaa16}.
A relation between massive IIA/IIB supergravities has also been analyzed, and it has been demonstrated that
  a duality exists between such  massive supergravities \cite{iiaa17}-\cite{iiaa18}. Thus, it is possible to study
  massive supergravity in string theory, and so massive gravity is also important in string theory.
  It may be noted that other solutions motivated by string theory has been also studied in massive gravity.
  In fact,   a brane in warped AdS spacetime has been constructed  in massive gravity \cite{mg1042}.
This was done by analyzing the effect of the mass term for the graviton on a infrared brane.
A nonextremal brane has also been analyzed in massive gravity \cite{mg1046}.
As there is a good motivation to both study
the massive gravity and gravity's rainbow from string theory,  it is both interesting and important
to study the rainbow deformation of massive gravity. So,  the target space metric flows and becomes
energy dependent, even in massive supergravities, and the bosonic part of this theory will be described
by massive gravity's rainbow. Thus, the   rainbow deformation of interesting solution massive gravity
has been recently studied  \cite{grmass}-\cite{grmass12}.

It may be noted that massive gravity can also be phenomenologically motivated from
accelerated cosmic expansion \cite{super}-\cite{super5}.
Even though there are problems with the massive gravity,  these problems can be resolved using the
  the Vainshtein mechanism \cite{4}-\cite{5}.
  However, the    Vainshtein mechanism  produces the  Boulware-Deser ghosts \cite{6}.
  It is possible to resolve this with ghosts fields  by using the dRGT mechanism
  \cite{7}-\cite{14}.
  It is possible to have a well defined initial value formulation
for massive gravity. In fact,
initial value constraints for spherically symmetric deformations of flat space, in
such a massive theory of gravity have been studied \cite{cosm14}. It has been demonstrated that
even though   the energy can be negative and even unbounded from below in certain  sector
of the theory, there is a physical sector of the theory, in which the energy
is positive and the ghosts are suppressed, and that the theory is stable \cite{cosm14}.
 The negative energy sector remains   disjointed, and does not have any effect on the
 physical sector of this theory.
The initial values for cosmological solutions have also been studied in massive gravity
\cite{cosm12}.
So, the theory has well defined posed initial value formulation, and
can be used to analyze the effects of graviton  mass on various physical phenomena.
The cosmological solutions in massive gravity have also been used to
obtain an upper bound on the graviton mass   \cite{cosmo12}.
The open FRW universes have been also studied in massive gravity,
and   it has been possible to obtain
universes  with   standard curvature   and an effective cosmological constant,
  such a theory of massive gravity \cite{cosmo14}.
In fact, various different  solutions in massive gravity have been studied,
 and the effect of such a mass on the physics of various systems has been discussed
 \cite{phys1}-\cite{phys2}. So, massive gravity is a very important theory of modified gravity,
 and it is important to study different solutions in massive gravity.

 In fact, as both rainbow gravity, and massive gravity are  motivated from string theory and phenomenology,
 we will   analyze a solution in the
 rainbow deformation of the massive gravity   \cite{grmass}-\cite{grmass12}. We will study Vaidya  solutions
 in this theory of massive gravity's rainbow because    Vaidya spacetime has
 used to study interesting physical system \cite{r1}-\cite{r4}.
The  Vaidya spacetime in massive gravity has been
 constructed \cite{r10}, and the  AdS/CFT has  been used to interstage field theory dual to
 a Vaidya-AdS solutions in massive gravity \cite{r12}. The Vaidya spacetime is also important in string theory \cite{stri12}-\cite{stri14}.
As Vaidya soltion is important in string theory,
 and string theory can also be used to motivate a rainbow deformation of massive gravity, we
 we will study the the Vaidya spacetime,
   massive gravity's rainbow. It may be noted even though Vaidya solution has been studied in gravity's rainbow \cite{r5}
   it has not been studied in massive gravity's rainbow, and so such it is interesting to analyze the Vaidya solution
   in massive gravity's rainbow.

\section{The Massive Gravity's Rainbow}
In this section, we  study the time-dependent  black hole solution using Vaidya metric.
This metric will be made energy dependent using  the framework of massive gravity's rainbow \cite{grmass}-\cite{grmass12}.
The four dimensional action for such a massive theory of gravity, can be written as
 \begin{equation}\label{field}
\mathcal{I}=\int d^{4}x \sqrt{-g}\left[\mathcal{R}+ \mathcal{M}^{2} \sum_{i}^{4}c_{i}\mathcal{U}_{i}(g,f)+\mathcal{L}_{m} \right],
\end{equation}
where $\mathcal{M}$ is the mass parameter in the massive   gravity.  Here $f$ is the  reference metric,   $c_i$ are constants,
and $\mathcal{U}_i$ are symmetric
polynomials of the eigenvalues of the $d\times d$
matrix ${\mathcal{K}^{\mu}}_{\nu}=\sqrt{g^{\mu\alpha}f_{\alpha\nu}} $. These symmetric polynomials  can be written as
\begin{eqnarray}\label{Ui}
&&\mathcal{U}_{1}=[\mathcal{K}],\nonumber\\
&&\mathcal{U}_{2}=[\mathcal{K}]^2 -[\mathcal{K}^2],\nonumber\\
&&\mathcal{U}_{3}=[\mathcal{K}]^3 -3[\mathcal{K}][\mathcal{K}^2]  + 2[\mathcal{K}^3],\nonumber\\
&&\mathcal{U}_{4}=[\mathcal{K}]^4 -6[\mathcal{K}^2][\mathcal{K}]^2  + 8[\mathcal{K}^3][\mathcal{K}]+3[\mathcal{K}^2]^2
-6[\mathcal{K}^4].
\end{eqnarray}
The square root in $\mathcal{K}$ can be  defined using
${(\sqrt{A})^{\mu}}_{\nu}{(\sqrt{A})^{\nu}}_{\lambda}={A^{\mu}}_{\lambda}$ and $\mathcal{K}={\mathcal{K}^{\mu}}_{\mu}$.
Now  the equation of motion from  this action, can be written as
\begin{equation}\label{field}
G_{\mu\nu}+\mathcal{M}^{2}\chi_{\mu\nu}=T_{\mu\nu},
\end{equation}
where $G_{\mu\nu}$ is the Einstein tensor,  and $\chi _{\mu \nu }$ is given by
\begin{eqnarray}\label{chi}
&\chi _{\mu \nu } =&-\frac{c_{1}}{2}\left( \mathcal{U}_{1}g_{\mu \nu }-%
\mathcal{K}_{\mu \nu }\right) -\frac{c_{2}}{2}\left( \mathcal{U}_{2}g_{\mu
\nu }-2\mathcal{U}_{1}\mathcal{K}_{\mu \nu }+2\mathcal{K}_{\mu \nu
}^{2}\right)\nonumber\\
 &&-\frac{c_{3}}{2}(\mathcal{U}_{3}g_{\mu \nu }-3\mathcal{U}_{2}
\mathcal{K}_{\mu \nu }
+6\mathcal{U}_{1}\mathcal{K}_{\mu \nu }^{2}-6\mathcal{K}_{\mu \nu }^{3})\nonumber\\ &&-\frac{c_{4}}{2}(\mathcal{U}_{4}g_{\mu \nu }-4\mathcal{U}_{3}\mathcal{K}_{\mu
\nu }+12\mathcal{U}_{2}\mathcal{K}_{\mu \nu }^{2}-24\mathcal{U}_{1}\mathcal{K
}_{\mu \nu }^{3}+24\mathcal{K}_{\mu \nu }^{4}).
\end{eqnarray}
We will analyze a   spatial reference metric, in the basis $(t, r, \theta, \phi)$
\cite{reference}, for this theory of massive gravity. Thus, we can write
\begin{equation}\label{fmetric}
f_{\mu \nu }=diag(0, 0, c^2 h_{ij}).
\end{equation}
where $h_{ij}$ is two dimensional Euclidean metric and $c$ is a positive constant.
We will now write  the Vaidya metric for this massive theory,  deformed by gravity's rainbow.
So, we will analyze the rainbow deformation  of the Vaidya metric,  in the case of advanced time coordinate.
These rainbow  deformations of this metric can be expressed as  \cite{grmass}-\cite{grmass12}
\begin{equation}
ds^{2} =-\frac{1}{\mathcal{F}^{2}(E)}\left(1-\frac{m(t,r)}{r}   \right)dt^2
+\frac{2}{\mathcal{F}(E)\mathcal{G}(E)}dtdr+\frac{1}{\mathcal{G}^{2}(E)}r^2
d\Omega_{2}^2,
 \end{equation}
where $\mathcal{F}(E)$ and $\mathcal{G}(E)$ are known as the gravity's rainbow
functions. It may be noted that here $E = E_{s}/E_P$, where $E_{s}$ is the maximum energy that a probe in that system can take, and $E_p$
is the Planck energy. So,  as $E_s/E_p \to 0$, $\mathcal{F}(E) =
\mathcal{G}(E) =1$, and the general relativity is recovered  in the IR
limit of the theory~\cite{Galan:2004st}-\cite{Galan:2004st1}. These rainbow functions are motivated from various
theoretical and phenomenology considerations.
The results   from
  loop quantum gravity   and   $\kappa$-Minkowski
noncommutative spacetime, have been used to motivate the following rainbow
functions \cite{AmelinoCamelia:1996pj}-\cite{amerev}
\begin{eqnarray}
 {\mathcal{F}(E/E_{p})=1~~~~\mathrm{and}~~~~\mathcal{G}(E/E_{p})=\sqrt{1-a\left(\frac{E}{E_{p}}\right)^q}.}
\label{RainFunc}
\end{eqnarray}
The modified dispersion relation  with constant
velocity of light, has been used to motivate the following rainbow functions ~\cite{MagSmolin}
\begin{equation}\label{MDR2}
 {\mathcal{F}(E/E_{p})=\mathcal{G}(E/E_{p})=\frac{1}{1-aE/E_{p}},}
\end{equation}
 The  hard spectra from gamma-ray
burster's, has been used to motivate the following rainbow functions
~\cite{AmelinoCamelia:1997gz}
\begin{equation}\label{MDR3}
 {\mathcal{F}(E/E_{p})=\frac{e^{a E/E_{p}}-1}{a
E/E_{p}}~~~\mathrm{and}~~~\mathcal{G}(E/E_{p})=1.}
\end{equation}
The maximum energy of the  system depends on the physical systems being analyzed,
and for black holes, this energy is equal to the energy of a quantum particle near the horizon.
This is because such a particle can be viewed as a probe for the geometry of the black hole.
In fact,   we can use the  uncertainty principle,  $\Delta p \geq 1/\Delta x $, to obtain
a bound on the energy of such a particle. So, we can write
$E_s \geq 1/\Delta x $, where $\Delta x$ is the uncertainty in position of the particle
near the horizon, and it is equal to   the radius of the event horizon. Thus, the bound
on the energy for a black hole can be written as
\begin{equation}
 E_s \geq 1/{\Delta x} \approx 1/{r_+}.
\end{equation}
It may be noted as the black hole evaporates due to the Hawking radiation, its radius reduces,
and this changes the bound on this maximum energy. So, this energy is a dynamical function,
and thus rainbow functions are also dynamical. Even though, we do not need the explicit
dynamical behavior of rainbow functions, it is important to know that they are dynamical,
and so they cannot be gauged away by rescaling of the metric.

Now, we assume the total energy-momentum tensor of the field equation (\ref{field}),
can be expressed in  the following form
\begin{equation}
T_{\mu\nu}=T_{\mu\nu}^{(n)}+T_{\mu\nu}^{(m)},
\end{equation}
 {where $T_{\mu\nu}^{(n)}$ and $T_{\mu\nu}^{(m)}$
are the energy-momentum tensor for the Vaidya null radiation and
the energy-momentum tensor of the perfect fluid, respectively. They can be
defined as}
\begin{eqnarray}
&&T_{\mu\nu}^{(n)}=\sigma l_{\mu}l_{\nu},\nonumber\\
&&T_{\mu\nu}^{(m)}=(\rho +p)(l_{\mu}n_{\nu}+l_{\nu}n_{\mu})+pg_{\mu\nu},
\end{eqnarray}
where $\sigma$, $\rho$ and $p$ are null radiation density, energy density
and pressure of the perfect fluid, respectively. In this regard, $l_{\mu}$ and $n_{\mu}$
are linearly independent future pointing null vectors,
\begin{equation}
l_{\mu}=\left(\frac{1}{\mathcal{F}(E)},0,0,0\right)~~~~~\&~~~
n_{\mu}=\left(\frac{1}{2\mathcal{F}(E)}\left(1-\frac{m(t,r)}{r} \right),-\frac{1}{\mathcal{G}(E)},0,0  \right),
\end{equation}
satisfying the following conditions
\begin{equation}
l_{\mu}l^{\mu}=n_{\mu}n^{\mu}=0~~~~\&~~~~l_{\mu}n^{\mu}=-1.
\end{equation}
Therefore, the non-vanishing components of the total energy-momentum tensor can be written as
\begin{eqnarray}
 T_{00}=\frac{\sigma}{\mathcal{F}^{2}(E)}+\frac{\rho}{\mathcal{F}^{2}(E)}\left(1-\frac{m(t,r)}{r}\right),&&
 T_{01}=-\frac{\rho}{\mathcal{F}(E)\mathcal{G}(E)}, \nonumber \\
T_{22}=\frac{pr^2}{\mathcal{G}^2(E)},
&&T_{33}=\frac{pr^2 sin^2\theta}{\mathcal{G}^2(E)}.
\end{eqnarray}
Using the metric ansatz (\ref{fmetric}), we obtain
\begin{equation}\label{fmetric}
{\mathcal{K}^{\mu}}_{ \nu }=diag\left(0, 0, \frac{c\mathcal{G}(E)}{r},\frac{c\mathcal{G}(E)}{r}\right).
\end{equation}
Therefore, we find that
\begin{eqnarray}\label{K1}
&&({\mathcal{K}^2)^{\mu}}_{ \nu }={\mathcal{K}^{\mu}}_{ \alpha }{\mathcal{K}^{\alpha}}_{ \nu }=diag\left(0, 0, \frac{c^{2}\mathcal{G}^2(E)}{r^{2}},\frac{c^{2}\mathcal{G}^2(E)}{r^{2}}\right),\nonumber\\
&&({\mathcal{K}^3)^{\mu}}_{ \nu }={\mathcal{K}^{\mu}}_{ \alpha }{\mathcal{K}^{\alpha}}_{ \beta }{\mathcal{K}^{\beta}}_{ \nu }=diag\left(0, 0, \frac{c^{3}\mathcal{G}^3(E)}{r^{3}},\frac{c^{3}\mathcal{G}^3(E)}{r^{3}}\right),\nonumber\\
&&({\mathcal{K}^4)^{\mu}}_{ \nu }=
{\mathcal{K}^{\mu}}_{ \alpha }{\mathcal{K}^{\alpha}}_{ \beta }{\mathcal{K}^{\beta}}_{ \lambda }{\mathcal{K}^{\lambda}}_{ \nu }
=diag\left(0, 0, \frac{c^{4}\mathcal{G}^4(E)}{r^{4}},\frac{c^{4}\mathcal{G}^4(E)}{r^{4}}\right).
\end{eqnarray}
We also obtain the following  quantities
\begin{eqnarray}\label{K2}
 [\mathcal{K}]={\mathcal{K}^{\mu}}_{ \mu }=\frac{2c\mathcal{G}(E)}{r},
  &&[\mathcal{K}^2]=({\mathcal{K}^2)^{\mu}}_{ \mu }=\frac{2c^{2}\mathcal{G}^2(E)}{r^{2}},\nonumber \\ {[\mathcal{K}^3]}=({\mathcal{K}^3)^{\mu}}_{ \mu }=\frac{2c^{3}\mathcal{G}^3(E)}{r^{3}},
&&[\mathcal{K}^4]=({\mathcal{K}^4)^{\mu}}_{ \mu }=\frac{2c^{4}\mathcal{G}^4(E)}{r^{4}}.
\end{eqnarray}
Now, using the Eqs.  (\ref{K1}), (\ref{K2}), and Eq. (\ref{Ui}), we obtain
\begin{eqnarray}\label{U}
\mathcal{U}_{1} =\frac{2c \mathcal{G}(E)}{r}, &&
\mathcal{U}_{2} =\frac{2c^{2}\mathcal{G}^{2}(E)}{r^{2}}, \nonumber\\
\mathcal{U}_{3}=0, &&
\mathcal{U}_{4} =0.
\end{eqnarray}
Using the Eqs.  (\ref{fmetric}), (\ref{K1}), (\ref{K2}) and (\ref{U}), we can obtain the non-vanishing components of
the massive gravity term $\chi_{\mu\nu}$ in the field equation
(\ref{field}) as
\begin{eqnarray}
&&\chi_{00}=\left[\frac{c_{1}c ~\mathcal{G}(E)}{r \mathcal{F}^{2}(E)}+ \frac{c_{2}c^2 ~\mathcal{G}^{2}(E)}{r^{2} \mathcal{F}^{2}(E)}\right]\left( 1-\frac{m}{r} \right) , \nonumber\\
&&\chi_{01}=\chi_{10}=-\frac{1}{r\mathcal{F}(E)}\left(c_{1}c+\frac{c_{2}c^2
~\mathcal{G}(E)}{r}\right), \nonumber\\
&&\chi_{22}=-\frac{c_{1}c~r}{2\mathcal{G}(E)}, \nonumber\\
&&\chi_{33}=-\frac{c_{1}~c~r\,sin^{2}\theta}{2\mathcal{G}(E)}.
\end{eqnarray}
 Then, for the ${00}$ component of the field equation (\ref{field}), we have
\begin{eqnarray}
&&\frac{\mathcal{G}(E)}{r^{3}}\left[ r\,\dot m \,\mathcal{F}(E)+r\,\mathcal{G}(E) \, m^{\prime} -\mathcal{G}(E)\,m\, m^{\prime} \right]\nonumber\\
&&=\sigma+\rho\left(1-\frac{m}{r}\right)-\mathcal{M}^{2}\left[\frac{c_{1}c \mathcal{G}(E)}{r }+ \frac{c_{2}c^2 \mathcal{G}^{2}(E)}{r^{2} }\right]\left( 1-\frac{m}{r} \right),
\end{eqnarray}
where dot and prime signs denote the derivative with respect to time and
radial coordinates, respectively. For the $01$ and $10$ component, we have
\begin{eqnarray}
-\frac{\mathcal{G}(E)m^{\prime}}{r^{2}}=-\frac{\rho}{\mathcal{G}(E)}+\frac{\mathcal{M}^{2}}{r}\left(c_{1}c+\frac{c_{2}c^2
~\mathcal{G}(E)}{r}\right).
\end{eqnarray}
Finally, for the $22$ and $33$ component, we obtain
\begin{equation}
-\frac{1}{2}r m^{\prime\prime }=\frac{pr^2}{\mathcal{G}^2(E)}+\frac{\mathcal{M}^{2}c_{1}c~r}{2\mathcal{G}(E)}.
\end{equation}
Thus, we have been able to analyze the Einstein equation in gravity's rainbow. In the next section,
we will analyze Vaidya spacetime in this massive gravity's rainbow.

 \section{Dynamics of the Collapsing System}
In this section,    we will first find a solution for the  field
equations describing this model. Then, we will analyze the dynamics of a collapsing system.
The  matter field will be assumed to follow a  barotropic equation of state, which is given by
\begin{equation}
p=k\rho,
\end{equation}
where $k$ is the barotropic parameter.
Now we can use  the Eqs. (18),(19) and (20),  and obtain an equation describing the
behavior of  $m(t,r)$  for this system,
\begin{equation}
r^{2}m^{\prime\prime}+6km+\frac{\left(1+3k\right)\mathcal{M}^{2}c_{1}c~r}{\mathcal{G}(E)}+6kc_{2}c^{2}\mathcal{M}^{2}r-6kf_{1}(t)=0,
\end{equation}
where $f_{1}(t)$ is an arbitrary function of time. This differential equation can be solved to
obtain  a solution for $m(t, r)$,
\begin{equation}
m(t,r)=f_{2}(t)r^{\omega_{1}}+f_{3}(t)r^{\omega_{2}}-\frac{\mathcal{M}^{2}c_{1}c\left(1+3k\right)r}
{\left(2-\omega_{1}\right)\left(2-\omega_{2}\right)\mathcal{G}(E)}-c_{2}c^{2}\mathcal{M}^{2}r+f_{1}(t),
\end{equation}
where $\omega_{1}=\frac{1}{2}\left(1+\sqrt{1-24k}\right)$,~
$\omega_{2}=\frac{1}{2}\left(1-\sqrt{1-24k}\right)$.
Here, $f_{2}(t)$ and $f_{3}(t)$ are arbitrary functions of time $t$. So, from  these equations,  we obtain the
admissible range of $k$, which is  $(-\infty, 1/24]$. Thus, the
metric given in Eq. (6),  can be expressed as
\begin{eqnarray}
 ds^{2}&=&\frac{1}{\mathcal{F}^{2}(E)}\left(-1+f_{2}(t)r^{\omega_{1}-1}+f_{3}(t)r^{\omega_{2}-1}-\frac{\mathcal{M}^{2}c_{1}c\left(1+3k\right)}
{\left(2-\omega_{1}\right)\left(2-\omega_{2}\right)\mathcal{G}(E)}-c_{2}c^{2}\mathcal{M}^{2}+\frac{f_{1}(t)}{r}\right)dt^2 \nonumber \\ &&
+\frac{2dtdr}{\mathcal{F}(E)\mathcal{G}(E)}+\frac{1}{\mathcal{G}^{2}(E)}r^2
d\Omega_{2}^2~.
\end{eqnarray}
This metric is  the generalized Vaidya metric in Massive
gravity's rainbow.

 {In this generalized Vaidya spacetime, the singularity can be either  a naked singularity or a black hole.
 The nature of this singularity is  determined by   the existence of outgoing radial null geodesics, which end  in the
past central singularity at $r=0$. Such geodesics exist for a
locally naked singularity, and do not exist for a black hole. So,
in massive gravity's rainbow, the singularity formed from
gravitational collapse can be either a naked singularity  or a
black hole. In general relativity, the cosmic censorship
hypothesis states that the gravitational singularity must
necessarily be covered by an event horizon. So, according to
cosmic censorship hypothesis only black hole can form from a
collapsing system. However, it has been demonstrated that
inhomogeneous dust cloud may form  a naked singularity
\cite{Eardley1}.  Interesting results have also been obtained by
studying fluid whose  equation of state is different from the
equation of state of dust \cite{Joshi1}. So, it is possible to
generalize the cosmic censorship hypothesis  \cite{Joshi2}.} As we
have to investigate the nature of  singularities in massive
gravity's rainbow, we can use such a  generalization of the cosmic
censorship hypothesis.

 { As this system is described by a time-dependent geometry,  the radius of shell at $r$, will also be a function of time $t$.
 We will describe such a radius by $R(t, ~r )$.   This system starts from
an initial time   $t=0$, and at that time, we have  $R(0,~r)=r$. It may be noted that for a
 inhomogeneous system,
different shells may become singular at different times. Now, for this system, we can have
future directed radial null geodesics coming out of the
singularity. These will have a  well defined tangent at  the singularity. So, for this system,
$\frac{dR}{dr}$ must tend to a finite limit, as the system
approach   the  past singularity.}
 {It is possible for the  system to reach  the points $(t_0, ~r)=0$. At this point,  the
singularity $R(t_0,0)=0$ occurs and  the matter shells are crushed to  a zero radius.
This  singularity at $r=0$,  is called a  central
singularity.}

{Now a  naked singularity  will form in this system, if   future
directed curves end  in the past singularity. So, for such a system,  the outgoing null geodesics
will end  in the past  central singularity, which is
at $r=0$ and $ t=t_0$. At such a point, $R(t_0, 0)=0$, and so for  these
geodesics,  we  have  $R\rightarrow 0$ as
$r\rightarrow 0$ \cite{Singh1}.}
{The equation for these outgoing radial null geodesics
can be obtained from the Eq. (6). Thus,   by putting $ds^{2}=0$ and
$d\Omega_{2}^{2}=0$}, we obtain
\begin{equation}
\frac{dt}{dr}=\frac{2\mathcal{F}(E)}{\mathcal{G}(E)\left(1-\frac{m(t,r)}{r}\right)}.
\end{equation}
Here $r=0,~t=0$ corresponds to a singularity in
this equation. Now if $X=\frac{t}{r}$, then
we can analyze  the limiting behavior of $X$,  as the system approaches  $r=0,~t=0$.
So,  if this limiting value  of $X$ is denoted by $X_{0}$, then we can write
\begin{eqnarray}
X_{0}
=\lim_{t\to0}\lim_{r\to0}~~ X
= \lim_{t\to0}\lim_{r\to0}\frac{t}{r}
= \lim_{t\to0}\lim_{r\to0} \frac{dt}{dr} \nonumber \\
= \lim_{t\to0}\lim_{r\to0}\frac{2\mathcal{F}(E)}{\mathcal{G}(E)\left(1-\frac{m(t,r)}{r}\right)}.\end{eqnarray}
We also use Eqs.  (25) and (28), and obtain
\begin{eqnarray*}
\frac{2}{X_{0}}&=& \lim_{t\to0}\lim_{r\to0}
\frac{\mathcal{G}(E)}{\mathcal{F}(E)}\left[1-f_{2}(t)r^{\omega_{1}-1}-f_{3}(t)
r^{\omega_{2}-1}\right. \nonumber \\ &&\left. +\frac{\mathcal{M}^{2}c_{1}c\left(1+3k\right)}{\left(1+\omega_{1}\right)\left(1+\omega_{2}\right)
\mathcal{G}(E)}  +c_{2}c^{2}\mathcal{M}^{2}-\frac{f_{1}(t)}{r}\right].
\end{eqnarray*}
Now, choosing $f_{1}(t)=\gamma t$,~~~$f_{2}(t)=\alpha
t^{1-\omega_{1}}$~~ and ~$f_{3}(t)=\beta t^{1-\omega_{2}}$, we
obtain the algebraic equation for $X_{0}$, which can  be written as
\begin{equation}
\alpha X_{0}^{1+\omega_{2}}+\beta X_{0}^{1+\omega_{1}}+\gamma
X_{0}^{2}-\left(1+c_{2}c^{2}\mathcal{M}^{2}\right)X_{0}-\frac{\left(1+3k\right)
\mathcal{M}^{2}c_{1}c}{\left(2-\omega_{1}\right)\left(2-\omega_{2}\right)\mathcal{G}(E)}+\frac{2\mathcal{F}(E)}{\mathcal{G}(E)}=0,
\end{equation}
where $\alpha$, $\beta$ and $\gamma$ are constants. If we only obtain
the non-positive solution of the equation, then a black hole will form in this system. However,  a naked singularity  can form
for positive roots of this equation.   Since this equation is  highly complicated,
it is extremely difficult to find out an analytic solution of
$X_{0}$. So, we will use numerical methods to find a   numerical solutions of $X_{0}$. This will be done by assigning
particular numerical values to the associated variables. In fact, as a specific rainbow function has been well motivated
\cite{AmelinoCamelia:1997gz, AmelinoCamelia:1999wk},   we will use this rainbow functions for analyzing this system,
\begin{equation}
\mathcal{F}(E)=1,~~~~~\mathcal{G}(E)=\sqrt{1-\eta\left(\frac{E_{1}}{E_{p}}\right)}
\end{equation}
In the above expressions, $E_{p}$ is the planck energy, given by
$E_{p}=1/\sqrt{G}=1.221 \times 10^{19}$ GeV, where $G$ is the
gravitational constant and $E_{1}=1.42 \times 10^{-13}$
\cite{AmelinoCamelia:1997gz, AmelinoCamelia:1999wk}. The
  value of $\eta$ has been estimated to be $\eta\approx 1$ \cite{AmelinoCamelia:1997gz}, and so in our study,  we will use
$\eta=1$.

\section{Conclusion and Discussion}
Now, we will  comment on  the numerical results obtained in this
paper. In Fig. 1, the contours $k-X_{0}$  were obtained for
different numerical values of $\alpha$, where other parameters were
fixed, in the massive gravity's rainbow.  {The admissible
plot range for the equation of state parameter, $k$ is $(-\infty,
1/24]$. In the figure, the plot range for $k$ has been taken as
$-2<k<1/24$ (from late to early universe). We observe that the
trajectories for different values of $\alpha$ almost coincide with
each other from $k=-2$ till around $k=-1/3$, i.e., the quintessence
and phantom regime (dark energy). But for $k>-1/3$, we observe that
this coincidence gradually disappears, and the red line
($\alpha=0.5$) diverges. However, this does not change the physics of the system much,
as all the trajectories remain in the positive
level of $X_0$. So,  the singularity formed is  a
naked singularity. At around $k=0$, the separation of the trajectories
becomes more pronounced. For greater values of $\alpha$ (blue
line), we see that there is a decreased tendency of formation of
naked singularity compared to lower values of $\alpha$. Even we
see that the blue line starts to take a dip around $k=0$.
The increased significance of $\alpha$
directly reflects on the function $f_{2}(t)$. Physically $k=0$
represents the dust regime and $k>0$ corresponds to early
universe. So, the $\alpha$ dependence of the system  will be more significant in the
earlier than in the later stages of the evolution of the universe. This is because in massive gravity's rainbow, the
spacetime is  energy dependent, and the energy in the earlier stages of the evolution of the universe is more than the energy
at the later stages of the evolution of the universe.
So, the rainbow functions are more important in the
physics of the early stages of the universe.   This is
the reasons that the   significance of $\alpha$ decreases at the later stages of the evolution of the
 universe.}

In Fig. 2, similar figures are obtained for different values of
$\beta$, where the other parameters are fixed. We observe that as
the values of $\beta$ increase, the trajectories push downwards
towards the $k$-axis. This indicate an increase in the tendency to
form black holes. However, for both figs. 1 and 2, it is clear
that the trajectories remain in the positive $X_{0}$ region, and a naked singularity forms from the
collapse of this system. In Figs. 3 and 4, the $k-X_{0}$ plots are obtained for
different values of $\gamma$ and $\mathcal{M}$, respectively.  These  plots also
indicate that a naked singularity is formed from the  collapse of this system.
We can observe from Fig. 3,  the tendency to form a  black
hole  increases with increase in the value of $\gamma$. We can also observe from Fig. 4, the tendency to form
a  black hole decreases with the increase in the value of  $\mathcal{M}$. So, the system can form
a naked singularity by increasing the  value of $\mathcal{M}$, and decreasing the value of $\gamma$.

In Figs. 5 and 6, we compare the $k-X_{0}$ contours of both massive
gravity and massive gravity's rainbow. In Fig. 5, the trajectories
are for different values of $\alpha$.  {From the
plot, we can observe that  for massive gravity's rainbow,  $\alpha$ does not play an important role in the
collapsing system, when $k<-1/3$. However,  for pure massive gravity
$\alpha$, does not play an important role throughout the domain.} Besides, in gravity's rainbow,
the tendency to form  black holes is greater than
that in pure massive gravity. In Fig. 6, similar plots are obtained
for different values of $\beta$. Here, it is also confirmed that
in gravity's rainbow, there is a greater tendency to form a
black hole. The above observation is again established in Figs. 7
and 8, where similar plots are generated by varying $\gamma$ and
$\mathcal{M}$, respectively. {Finally we observe that
in all the figures, there are small portions of lines which
nearly vanish around the $k$-axis, for small values of $k$.
As this system was very  complicated, we could not find
an analytical solution for Eq. (30). So, we
obtained  numerical solution for this equation,  using particular values for the
parameters. The
vanishing lines in the $k-X_0$ plane, are produced from the   noise in
the numerical solution, and do not have physical significance.}

\begin{figure}[ht]
\includegraphics[height=3in,width=3in]{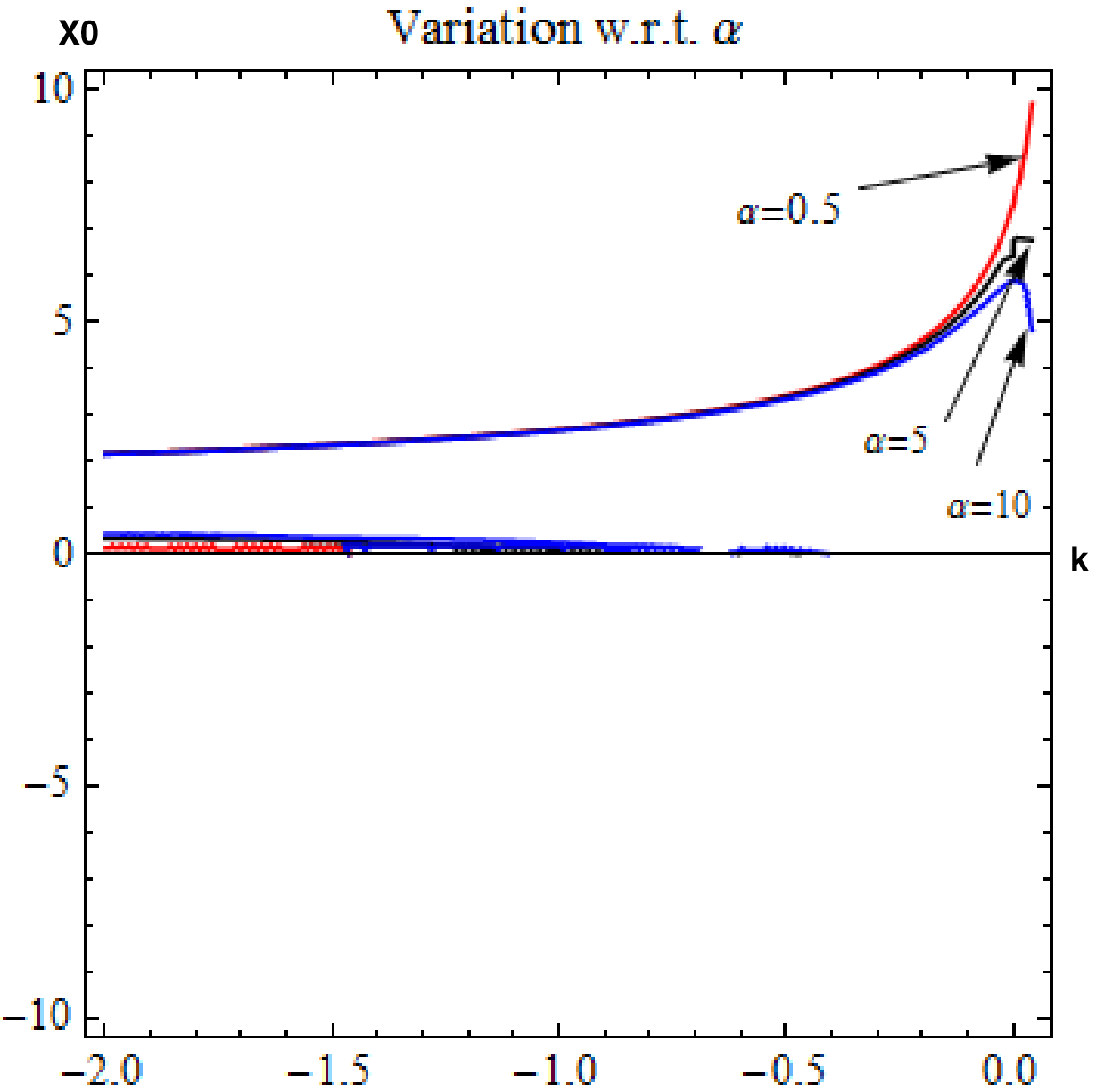}~~~~~~~\includegraphics[height=3in,width=3in]{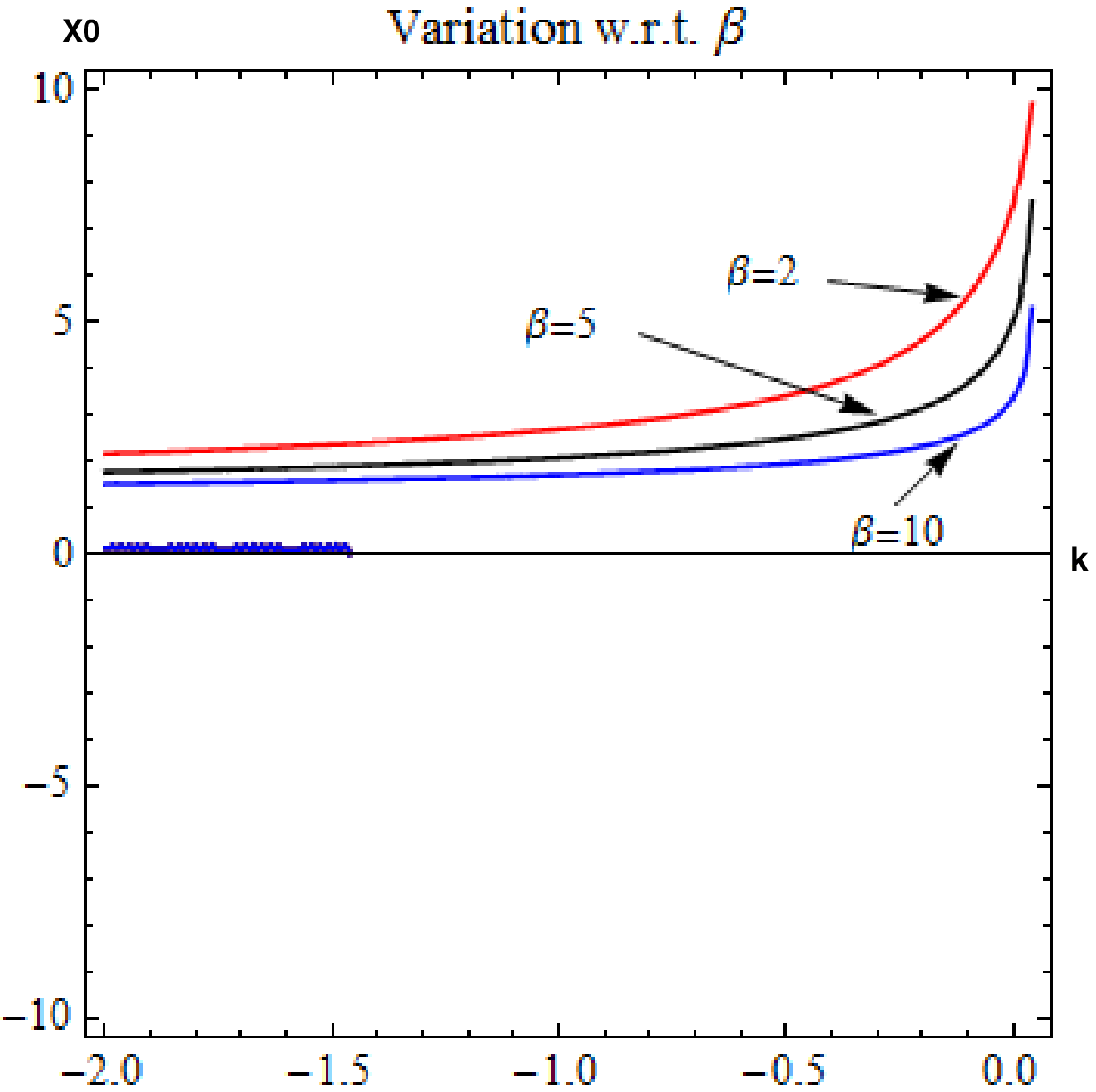}~~~~~~~\\\\

~~~~~~~~~~~~~~~~~~~~~~~~~Fig.1~~~~~~~~~~~~~~~~~~~~~~~~~~~~~~~~~~~~~~~~~~~~~~~~~~~~~~~Fig.2~~~~~~~~~\\

\vspace{2mm} \textit{\textbf{Figs 1 and 2} show the variation of
$X_{0}$ with $k$ for different values of $\alpha$ and $\beta$
respectively in massive gravity's rainbow.\\\\ In Fig.1 the other
parameters are fixed at $\beta=2$, $\gamma=3$, $c=0.8$, $c_{1}=4$,
$c_{2}=2$, $\mathcal{M}=5$, $\eta=1$, $E_{1}=1.42 \times
10^{-13}$, $E_{p}=1.221 \times 10^{19}$.\\\\ In Fig.2 the other
parameters are taken as $\alpha=0.5$, $\gamma=3$, $c=0.8$,
$c_{1}=4$, $c_{2}=2$, $\mathcal{M}=5$, $\eta=1$, $E_{1}=1.42
\times 10^{-13}$, $E_{p}=1.221 \times 10^{19}$.}
\end{figure}

\begin{figure}[ht]
\includegraphics[height=3in, width=3in]{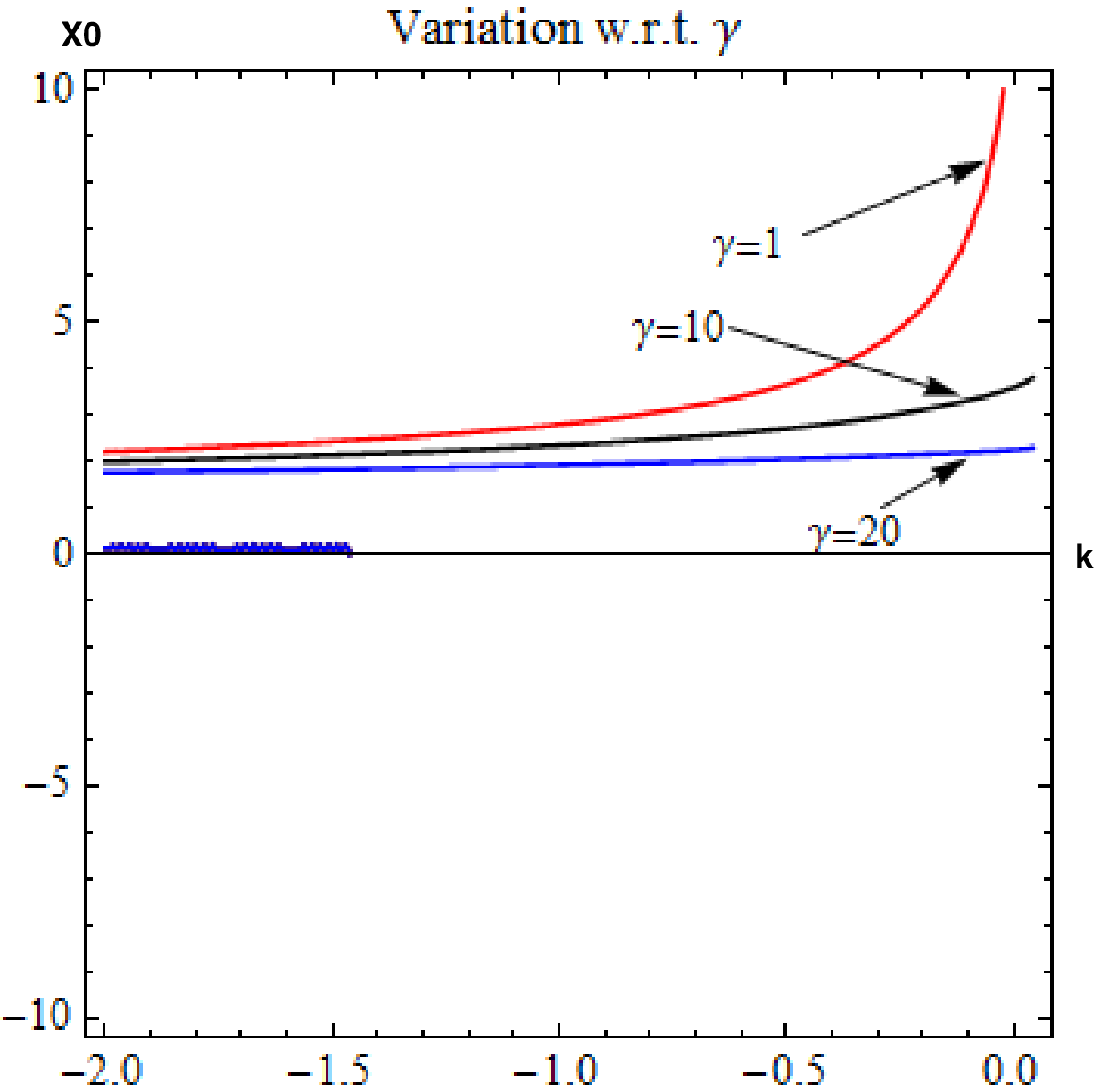}~~~~~~~\includegraphics[height=3in,width=3in]{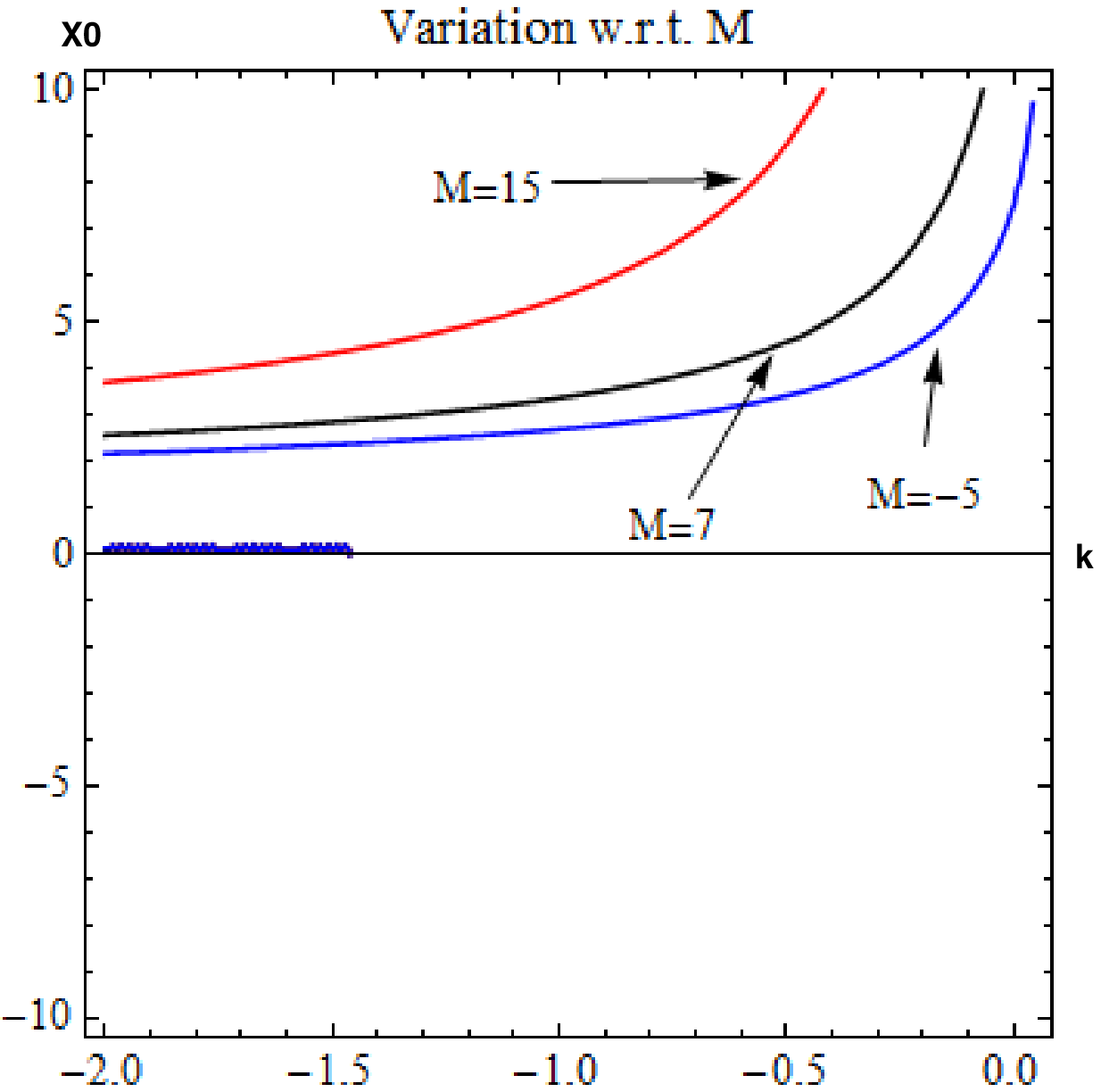}~~~~~~~\\\\

~~~~~~~~~~~~~~~~~~~~~~~~~Fig.3~~~~~~~~~~~~~~~~~~~~~~~~~~~~~~~~~~~~~~~~~~~~~~~~~~~~~~~~~~Fig.4~~~~~~~~\\

\vspace{2mm} \textit{\textbf{Figs 3 and 4} show the variation of
$X_{0}$ with $k$ for different values of $\gamma$ and
$\mathcal{M}$ respectively in massive gravity's rainbow.\\\\ In
Fig.3 the other parameters are fixed at $\alpha=0.5$, $\beta=2$,
$c=0.8$, $c_{1}=4$, $c_{2}=2$, $\mathcal{M}=5$, $\eta=1$,
$E_{1}=1.42 \times 10^{-13}$, $E_{p}=1.221 \times 10^{19}$.\\\\ In
Fig.4 the other parameters are taken as $\alpha=0.5$, $\beta=2$,
$\gamma=3$, $c=0.8$, $c_{1}=4$, $c_{2}=2$, $\eta=1$, $E_{1}=1.42
\times 10^{-13}$, $E_{p}=1.221 \times 10^{19}$.}
\end{figure}

\begin{figure}[ht]
\includegraphics[height=3in, width=3in]{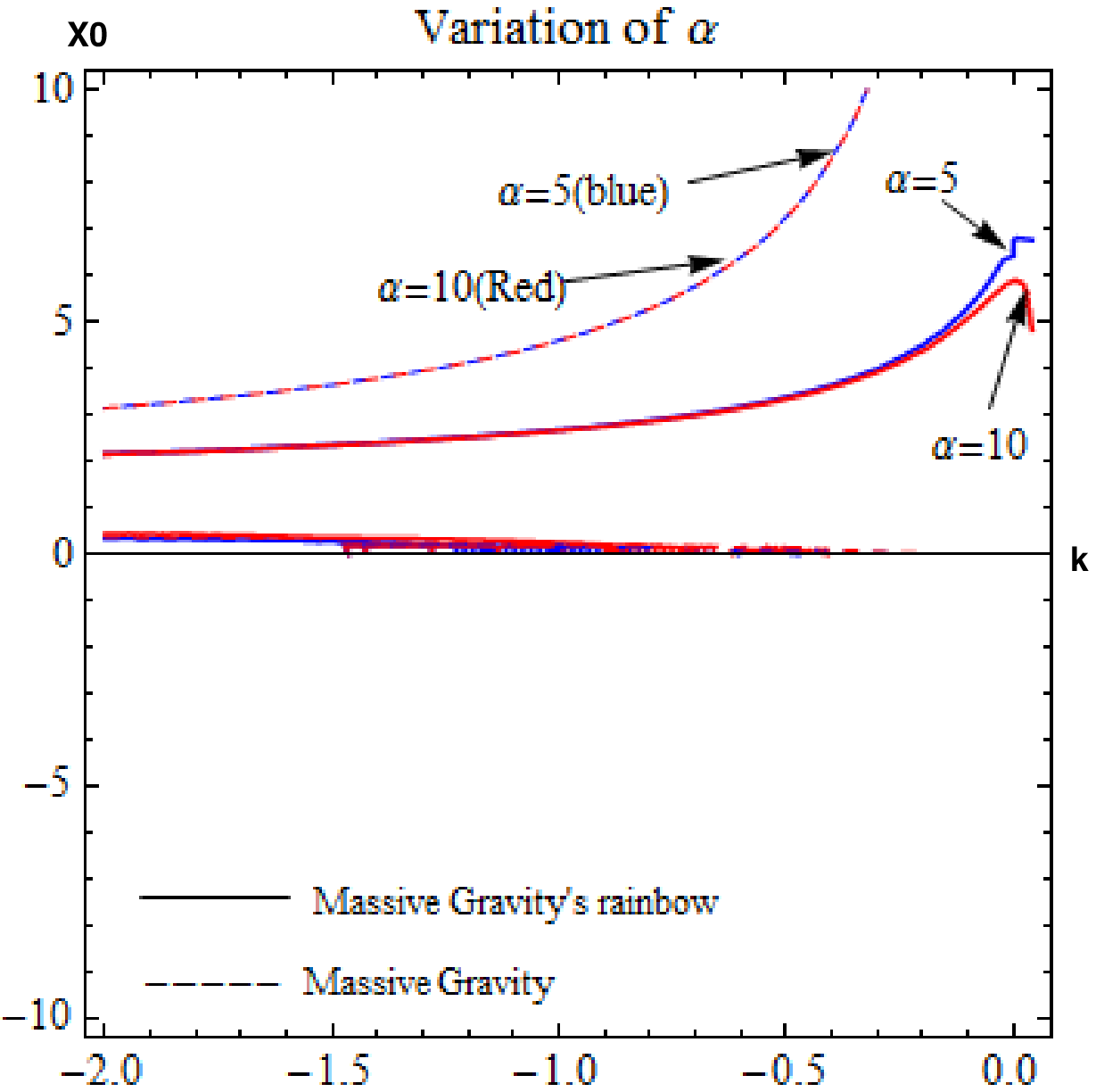}~~~~~~~\includegraphics[height=3in,width=3in]{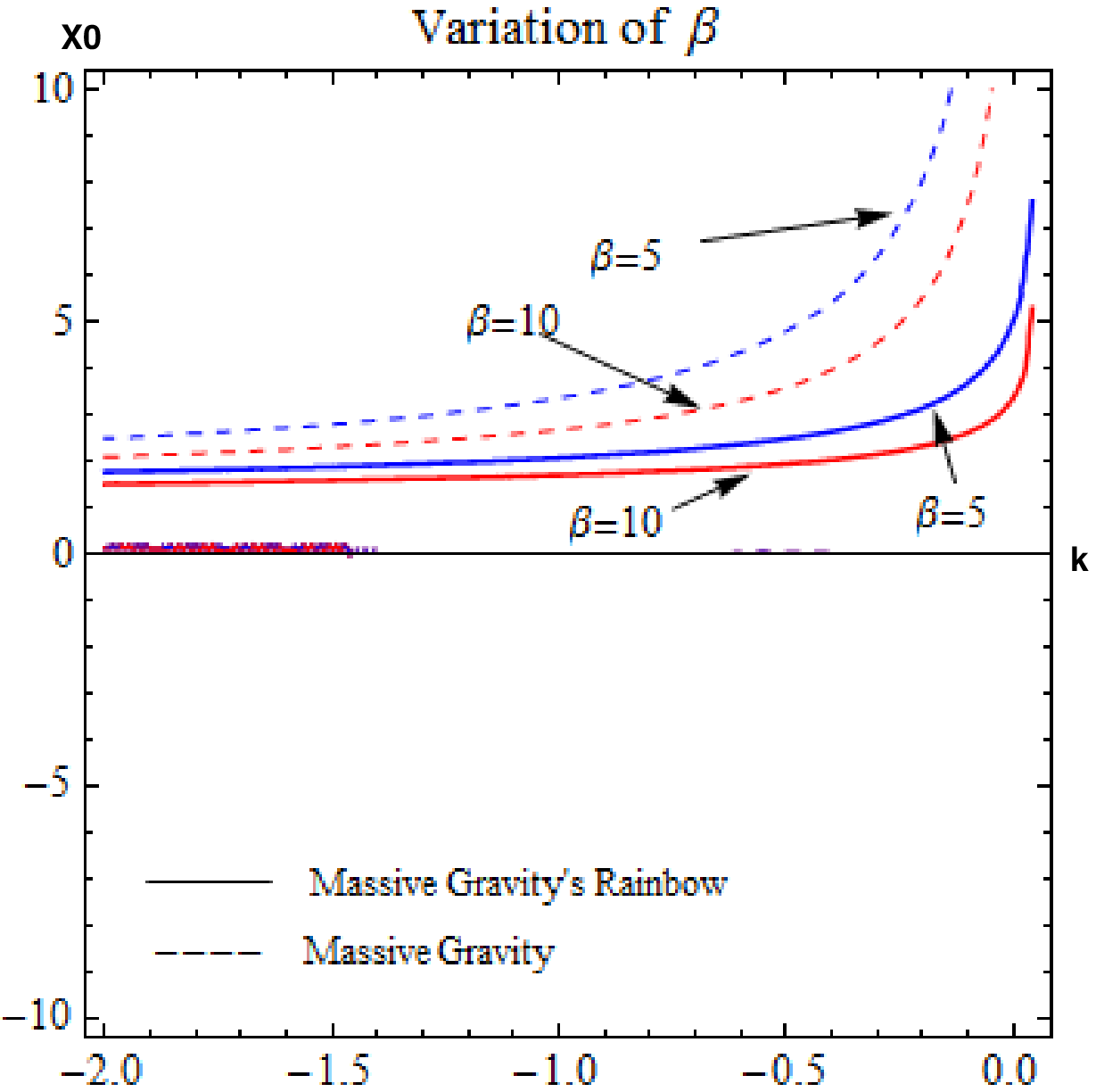}~~~~~~~\\\\

~~~~~~~~~~~~~~~~~~~~~~~~~~Fig.5~~~~~~~~~~~~~~~~~~~~~~~~~~~~~~~~~~~~~~~~~~~~~~~~~~~~~~~Fig.6~~~~~~\\

\vspace{2mm} \textit{\textbf{Figs 5 and 6} show the variation of
$X_{0}$ with $k$ for different values of $\alpha$ and $\beta$
respectively in a comparative scenario between Massive gravity and Massive gravity's rainbow.\\\\
In Fig.5 the other parameters are fixed at $\beta=2$, $\gamma=3$,
$c=0.8$, $c_{1}=4$, $c_{2}=2$, $\mathcal{M}=5$, $\eta=1$,
$E_{1}=1.42 \times 10^{-13}$, $E_{p}=1.221 \times 10^{19}$.\\\\ In
Fig.6 the other parameters are taken as $\alpha=0.5$, $\gamma=3$,
$c=0.8$, $c_{1}=4$, $c_{2}=2$, $\mathcal{M}=5$, $\eta=1$,
$E_{1}=1.42 \times 10^{-13}$, $E_{p}=1.221 \times 10^{19}$.}

\end{figure}

\begin{figure}[ht]
\includegraphics[height=3in, width=3in]{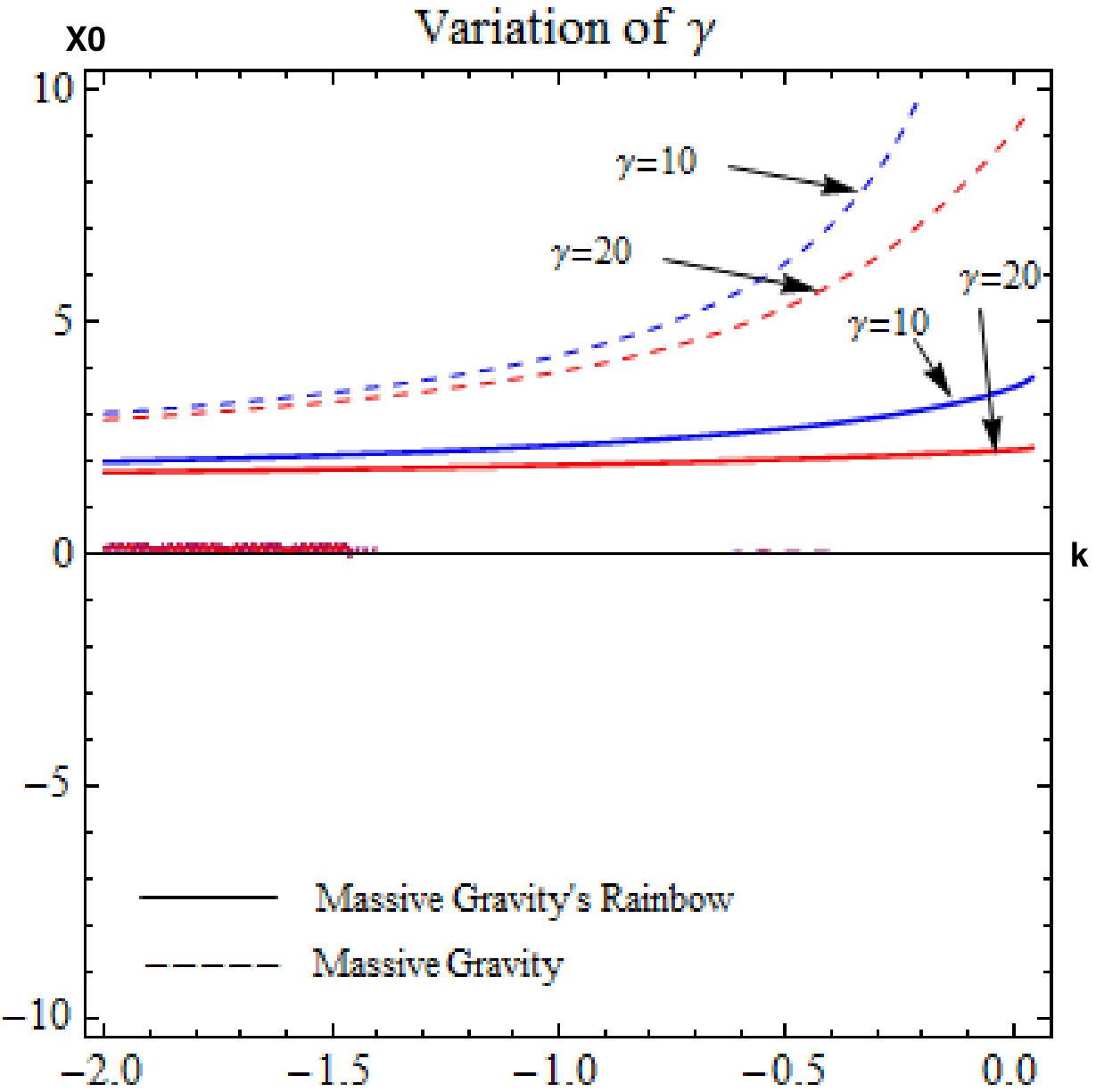}~~~~~~~\includegraphics[height=3in,width=3in]{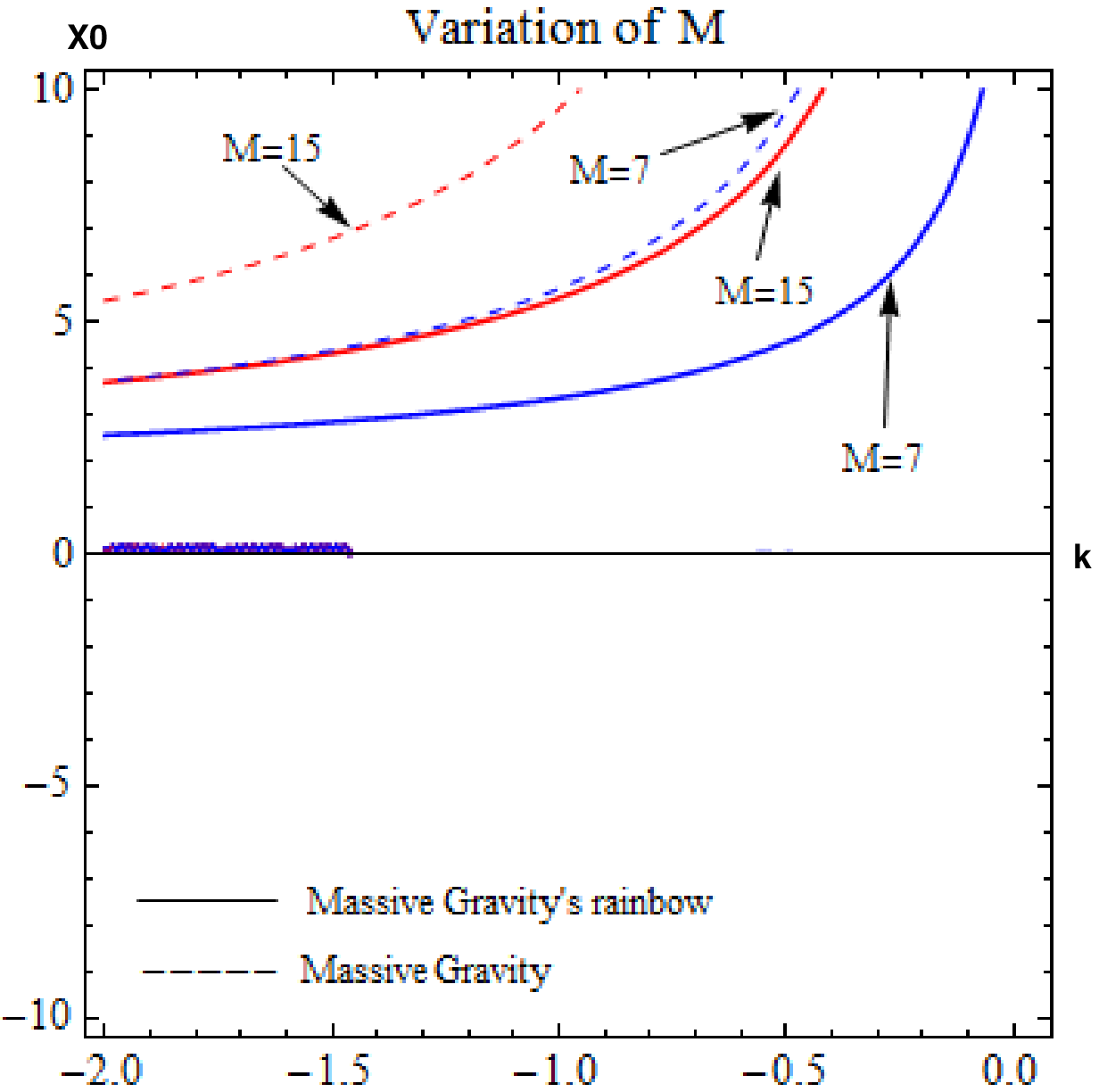}~~~~~~~\\\\

~~~~~~~~~~~~~~~~~~~~~~~~~~~Fig.7~~~~~~~~~~~~~~~~~~~~~~~~~~~~~~~~~~~~~~~~~~~~~~~~~~~~~~~Fig.8~~~~~~\\

\vspace{2mm} \textit{\textbf{Figs 7 and 8} show the variation of
$X_{0}$ with $k$ for different values of $\gamma$ and
$\mathcal{M}$ respectively in a comparative scenario between Massive gravity and Massive gravity's rainbow.\\\\
In Fig.7 the other parameters are fixed at $\alpha=0.5$,
$\beta=2$, $c=0.8$, $c_{1}=4$, $c_{2}=2$, $\mathcal{M}=5$,
$\eta=1$, $E_{1}=1.42 \times 10^{-13}$, $E_{p}=1.221 \times
10^{19}$.\\\\ In Fig.8 the other parameters are taken as
$\alpha=0.5$, $\beta=2$, $\gamma=3$, $c=0.8$, $c_{1}=4$,
$c_{2}=2$, $\eta=1$, $E_{1}=1.42 \times 10^{-13}$, $E_{p}=1.221
\times 10^{19}$.}
\end{figure}

In this paper, we have constructed a theory of massive gravity's
rainbow. This was done by analysing the energy dependent
deformation of massive gravity. In the construction of massive
gravity, we have used the
Vainshtein mechanism and the dRGT mechanism. Then, this theory has been deformed by rainbow
functions. We have analyzed  radiating Vaidya black hole solution
in this theory of massive gravity's rainbow. The effects of both
the graviton mass and rainbow deformation have been studied for
a time-dependent system. It may be noted that the AdS solution in massive
gravity, and the AdS/CFT correspondence corresponding to this AdS
solution have been studied \cite{ads}-\cite{cft}. In fact, the
holographic entanglement entropy  for  massive gravity has also
been studied \cite{ee}, and it has been demonstrated that  for
such systems both  first order and  second order phase transitions
can occur. The holographic complexity for massive gravity has also
been studied  \cite{cc}. This holographic complexity of a boundary
theory is dual to a volume in the bulk, just as the holographic
entanglement entropy is dual to an area in the bulk. It would be
interesting to study the rainbow deformation of such solutions.
This can be done by making the bulk metric to depend on the energy of
the probe. Then, deformation of the bulk metric can be done using
suitable rainbow functions. It would be interesting to investigate
the holographic entanglement entropy and holographic complexity of
massive gravity deformed by suitable rainbow functions.

\section*{Acknowledgements}

P. Rudra acknowledges University Grants Commission, Govt. of India for providing research project grant (No. F.PSW-061/15-16 (ERO)). F. Darabi acknowledges financial support of Azarbaijan Shahid Madani University (No. S/5749-ASMU ) for the Sabbatical Leave, and thanks the hospitality of ICTP (Trieste) for providing support during the Sabbatical Leave.\\


\end{document}